\documentclass[11pt,a4paper]{article}
\usepackage{graphicx}
\usepackage{amssymb}
\usepackage{amsmath}\usepackage{braket}
\usepackage{listings}
\usepackage{subfig}
\usepackage{amssymb}
\usepackage{cases}
\usepackage{comment}
\usepackage{ltgtsim}
\usepackage{jcappub}
\newcommand{\lambdabar}{\lambda \kern -0.5em\raise 0.5ex \hbox{--}}
\newcommand{\dif}[2]{\frac{d #1}{d #2}}
\newcommand{\pdif}[2]{\frac{\partial #1}{\partial #2}}
\def\slashchar#1{\setbox0=\hbox{$#1$} 
\dimen0=\wd0 
\setbox1=\hbox{/} \dimen1=\wd1 
\ifdim\dimen0>\dimen1 
\rlap{\hbox to \dimen0{\hfil/\hfil}} 
#1 
\else 
\rlap{\hbox to \dimen1{\hfil$#1$\hfil}} 
/ 
\fi}

\title{A new algorithm for calculating the curvature perturbations in stochastic inflation}

\author[a,b]{Tomohiro Fujita,}
\author[a,c]{Masahiro Kawasaki,}
\author[a,b,d]{Yuichiro Tada}
\author[c]{and Tomohiro Takesako}
\affiliation[a]{Kavli Institute for the Physics and Mathematics of the
Universe (Kavli IPMU), TODIAS,  the University of Tokyo, 5-1-5
Kashiwanoha, Kashiwa, 277-8583, Japan}
\affiliation[b]{Department of Physics, the University of Tokyo, Bunkyo-ku
113-0033, Japan}
\affiliation[c]{Institute for Cosmic Ray Research, the University of Tokyo,
5-1-5 Kashiwa-no-Ha, Kashiwa, Chiba, 277-8582, Japan}
\affiliation[d]{Advanced Leading Graduate Course for Photon Science (ALPS), the University of Tokyo, Bunkyo-ku
113-0033, Japan}

\emailAdd{tomohiro.fujita@ipmu.jp}
\emailAdd{kawasaki@icrr.u-tokyo.ac.jp}
\emailAdd{yuichiro.tada@ipmu.jp}
\emailAdd{takesako@icrr.u-tokyo.ac.jp}

\abstract{We propose a new approach for calculating the curvature perturbations produced during inflation
in the stochastic formalism. In our formalism, the fluctuations of the e-foldings are directly calculated
without perturbatively expanding the inflaton field and they are connected to the curvature perturbations
by the $\delta N$ formalism. The result automatically includes the contributions of the higher order perturbations because 
we solve the equation of motion non-perturbatively. In this paper, we analytically prove that our result (the power spectrum and the 
nonlinearity parameter) is consistent with the standard result in single field slow-roll inflation. We also describe the algorithm for 
numerical calculations of the curvature perturbations in more general inflation models.}


\arxivnumber{1308.4754}

\begin{document}

\begin{flushright}
ICRR-Report-658-2013-7
\\
IPMU 13-0154
\end{flushright}

\maketitle
\flushbottom

\section{Introduction}
Inflation~\cite{Guth:1980zm,Starobinsky:1980te,Sato:1980yn,Linde:1981mu,Albrecht:1982wi} is a plausible model of the early 
universe and now believed by many physicists.
It solves various problems such as the horizon problem, the flatness problem, the monopole problem
and so on~\cite{Brout:1977ix,Starobinsky:1979ty,Kazanas:1980tx}. Furthermore, simple inflation models produce almost scale invariant 
curvature perturbations which are verified by the observations of the cosmic microwave background (CMB) or 
the large scale structure~\cite{Ade:2013ktc}. 

Inflation is presumably caused by one or more real scalar fields which are called the inflatons and
the curvature perturbations are originated from the quantum fluctuations of these scalar fields. In usual analysis of
the dynamics of inflation, the inflaton field is divided into the spatially homogeneous classical background part and the perturbative quantum 
fluctuations part. Then the contributions of the quantum fluctuations to the background dynamics
are often neglected. However, in some inflation models,
the quantum kicks to the background are essential. For example, it is possible that the background inflaton field not only rolls down but also
climbs up the potential due to the quantum kicks. 
In these cases, inflation will not end eternally
(which is called \emph{eternal inflation}~\cite{Vilenkin:1983xq,Linde:1986fd}). The \emph{stochastic formalism}~\cite{Starobinsky:1986fx,
Starobinsky:1994bd,Sasaki:1987gy,Nakao:1988yi,Nambu:1988je,Mollerach:1990zf,Martin:2011ib,Enqvist:2011pt,Kawasaki,
Finelli:2008zg,Finelli:2010sh,Sanchez:2012tk,Sanchez:2013zst}
can take into account the quantum effects by treating the quantum kicks as stochastic noise in the equation of motion
as we will see in section \ref{stochastic}. 

When the quantum kicks are taken into consideration, the background field is no longer homogeneous because the noise in each
Hubble scale region has independent values. In other words, each Hubble scale universe evolves separately. Therefore
a linear perturbation theory where the background field is assumed to be homogeneous
might not give correct curvature perturbations. Especially in a highly non-Gaussian case, for example, 
it is difficult to calculate the curvature perturbations in a perturbative theory because the contributions of the
higher order perturbations are crucial.

In this paper, we propose a new algorithm for calculating the curvature perturbations in the stochastic formalism.
In our algorithm, the fluctuations of the e-foldings are directly calculated
and they are connected to the curvature perturbations with use of the $\delta N$ formalism~\cite{Lyth:2004gb,
Starobinsky:1986fxa,Salopek:1990jq,Sasaki:1995aw,Sasaki:1998ug}. 
Since we do not need to expand the inflaton field perturbatively in our formalism,
results 
automatically include the contributions of the higher order perturbations of the inflaton field.
Therefore, when these contributions are important, such as highly non-Gaussian inflation models,
it becomes a quite useful method. It is also advantageous when we calculate the higher order quantities
such as the non-Gaussianity of the curvature perturbations. According to the recent observation of CMB~\cite{Ade:2013ktc}, the curvature perturbations
are so small and almost Gaussian at the CMB scale, but it is possible that they are large and/or quite non-Gaussian at smaller scales,
which leads to significant cosmological effect.
In fact, some interesting astronomical objects like primordial black holes
(PBHs)~\cite{Carr:2009jm,Carr:1975qj,Carr:1976zz,Carr:2003bj} and ultracompact minihalos
(UCMHs)~\cite{Bringmann:2011ut,Josan:2010vn,Li:2012qha}
can be formed when the curvature perturbations are large and non-Gaussian. Our non-perturbative formalism is needed in such cases.

If we consider a single field inflation model with sufficiently flat inflaton potential, the curvature perturbations
can be analytically calculated even in our formalism 
with the help of techniques of stochastic calculus.
Then we can compare the power
spectrum and the nonlinearity parameter of the curvature perturbations
calculated in our formalism with those in usual perturbation theory.
It is found the results are consistent and thus our formalism is verified
at least in single slow-roll inflation models.
Moreover, our formalism is applicable to wider class of inflation models
by using numerical calculations. We will describe a calculation algorithm
for more general inflation models.

The power spectrum of the curvature perturbations was firstly derived in the stochastic formalism by Kunze~\cite{Kunze:2006tu}.
Ref. \cite{Kunze:2006tu} obtained the power spectrum from the time derivative of the variance of the e-foldings, which we also adopt in our formalism.
However, in that article, the inflaton field is expanded perturbatively with respect to the noise term~\cite{Martin:2005ir,Martin:2005hb}
and the validity of this perturbative expansion is non-trivial. Furthermore, their result deviates from the standard result of the usual perturbation
theory even in single slow-roll inflation where the standard linear perturbation theory works well and gives precise results. Therefore,
the interpretation of this deviation is difficult. In contrast, in our formalism, the curvature perturbations can be calculated non-perturbatively and
the result is consistent with the standard formalism.\footnote{ 
During preparation of this paper, another approach to curvature perturbations in the stochastic formalism has been provided
in ref. \cite{Levasseur:2013ffa,Levasseur:2013tja} and they also seem to relate with our work.}

The structure of this paper as follows. In section \ref{stochastic}, we briefly review the stochastic formalism.
In section \ref{analytic}, we explain the techniques of stochastic calculus in the first place. Then we will show
how the power spectrum and the non-Gaussianity of the curvature perturbations can be calculated
and prove that the result in single slow-roll inflation is consistent with that of the standard formalism.
In section \ref{numerical}, the extension to more general inflation models is discussed.
Finally, section \ref{Conclusion} is devoted to conclusions of this paper.

\section{Stochastic inflation}\label{stochastic}

In the stochastic approach, we divide the real inflaton field $\phi$ into 
two parts, IR mode and UV mode, as
\begin{eqnarray}
        \phi(\mathbf{x},t)&=&\phi_\mathrm{IR}(\mathbf{x},t)+\phi_\mathrm{UV}(\mathbf{x},t), \\
        \phi_\mathrm{IR}(\mathbf{x},t)&=&\int\frac{d^3k}{(2\pi)^3}\theta(\epsilon a(t)H(t)-k)\phi_\mathbf{k}(t)
        e^{-i\mathbf{k}\cdot\mathbf{x}},\\
        \phi_\mathrm{UV}(\mathbf{x},t)&=&\int\frac{d^3k}{(2\pi)^3}(1-\theta(\epsilon a(t)H(t)-k))\phi_\mathbf{k}(t)
        e^{-i\mathbf{k}\cdot\mathbf{x}} , 
\end{eqnarray}
where $\phi_\mathbf{k}(t)$ is the Fourier mode of the inflaton field $\phi(\mathbf{x},t)$, 
$\theta(z)$ is the step function, $H(t)$ is the Hubble parameter, $a(t)$ is the scale factor 
and $\epsilon$ is a small constant parameter. In this paper, following the original 
works~\cite{Starobinsky:1986fx,Starobinsky:1994bd}, we choose the step function as the window function splitting the IR mode and 
the UV mode.\footnote{Note that a validity of a sharp cutoff window function is under discussion~\cite{Winitzki:1999ve,Casini:1998wr}.} 
We treat the IR mode as a classical stochastic background field and are interested only in this mode.  

Let us consider the following action for the inflaton field:
\begin{eqnarray}
        S=\int d^4x\sqrt{-g}\left(\frac{1}{2}\partial_\mu\phi\partial^\mu\phi-V(\phi)\right),
\end{eqnarray}
and then the equations of motion in the Hamiltonian formulation are given by
\begin{numcases}
	{}
	\pi=\dot{\phi}, \label{eom1-1} \\
	\dot{\pi}+3H\pi-a^{-2}\nabla^2\phi-V^\prime(\phi)=0, \label{eom1-2}
\end{numcases}
where a dot and a prime denote derivatives respect to the time and the inflaton field respectively.
In a way similar to the inflaton field, we also divide the conjugate field $\pi$ into the IR and UV modes as
\begin{eqnarray}
        \pi_\mathrm{IR}(\mathbf{x},t)&=&\int\frac{d^3k}{(2\pi)^3}\theta(\epsilon a(t)H(t)-k)\dot{\phi}_\mathbf{k}(t)
        e^{-i\mathbf{k}\cdot\mathbf{x}},  \\
        \pi_\mathrm{UV}(\mathbf{x},t)&=&\int\frac{d^3k}{(2\pi)^3}(1-\theta(\epsilon a(t)H(t)-k))\dot{\phi}_\mathbf{k}(t)
        e^{-i\mathbf{k}\cdot\mathbf{x}}.
\end{eqnarray}
It should be noted that $\dot{\phi}_\mathrm{IR}\neq\pi_\mathrm{IR}$ because of the time dependence of $\epsilon aH$.
Substituting these definitions into the equations of motion (\ref{eom1-1}) and (\ref{eom1-2}), with approximation,
\begin{eqnarray}
        V^\prime(\phi)\simeq V^\prime(\phi_\mathrm{IR})+V^{\prime\prime}(\phi_\mathrm{IR})\phi_\mathrm{UV},
\end{eqnarray}
and equation of motion for $\phi_\mathbf{k}$,
\begin{eqnarray}
        \ddot{\phi}_\mathbf{k}+3H\dot{\phi}_\mathbf{k}+\left(\frac{k^2}{a^2}+V^{\prime\prime}(\phi_\mathrm{IR})\right)\phi_\mathbf{k}=0,
\end{eqnarray}
we obtain the equations for the IR mode,
\begin{numcases}
	\displaystyle
	\dot{\phi}_\mathrm{IR}=\pi_\mathrm{IR}+\epsilon aH^2\int\frac{d^3k}{(2\pi^3)}\delta(\epsilon aH-k)
	\phi_\mathbf{k}e^{-i\mathbf{k}\cdot\mathbf{x}}, \label{eom2-1} \\
	 \displaystyle
	 \dot{\pi}_\mathrm{IR}=-3H\pi_\mathrm{IR}-V^\prime(\phi_\mathrm{IR})+\epsilon aH^2\int\frac{d^3k}{(2\pi^3)}\delta(\epsilon aH-k)
	 \pi_\mathbf{k}e^{-i\mathbf{k}\cdot\mathbf{x}}. \label{eom2-2}
\end{numcases}
Here, the spatial derivative of the IR mode is neglected. Note that the Dirac delta function comes from the time derivative of 
the step function.

From now on, we consider only one spatial point. Then we can fix the spatial point to the origin, $\mathbf{x}=0$,
without loss of generality. We denote the last terms of eqs.~(\ref{eom2-1}) and (\ref{eom2-2}) as $\xi_\phi(t)$ and $\xi_\pi(t)$ respectively.
Their expected values  vanish because $\phi_\mathbf{k}$ and $\pi_\mathbf{k}$ are vacuum fluctuations.
Furthermore, with $k_c(t)=\epsilon aH$, the expected value of $\xi_\phi(t)\xi_\phi(t^\prime)$
is given by
\begin{eqnarray}
        \braket{\xi_\phi(t)\xi_\phi(t^\prime)}&=&k_c(t)k_c(t^\prime)H^2\int\frac{d^3kd^3k^\prime}{(2\pi)^6}
        \delta(k_c(t)-k)\delta(k_c(t^\prime)-k^\prime)\braket{\phi_\mathbf{k}(t)\phi_\mathbf{k^\prime}(t^\prime)} \nonumber \\
        &=&k_c^2(t)H^2\int\frac{d^3k}{(2\pi)^3}\delta(k_c(t)-k)\frac{\delta(t-t^\prime)}{k_c(t)H}
        \frac{2\pi^2\mathcal{P}_\phi(t,k)}{k^3} \nonumber \\
        &=&H\mathcal{P}_\phi(t,k_c)\delta(t-t^\prime),
\label{2 point xi}
\end{eqnarray}
where
\begin{eqnarray}
        \mathcal{P}_\phi(t,k)\delta(\mathbf{k}+\mathbf{k^\prime})=\frac{k^3}{2\pi^2}\braket{\phi_\mathbf{k}\phi_\mathbf{k^\prime}}.
\end{eqnarray}
With $\epsilon\ll1$, $k_c=\epsilon aH$ corresponds to the scale much larger than the horizon and we suppose that $\mathcal{P}_\phi(t,k_c)\to(H/2\pi)^2$.\footnote{This is a massless case
spectrum. In standard single inflation, the inflaton is almost massless and this assumption is natural. However, in more general
cases, the amplitude of the noise might not be constant. In those cases, numerical calculations are needed. }
Since $\xi_\phi (t)$ satisfies
\begin{eqnarray}
        \braket{\xi_\phi(t)}=0,
        \qquad
        \braket{\xi_\phi(t)\xi_\phi(t^\prime)}=\frac{H^3}{(2\pi)^2}\delta(t-t^\prime),
\end{eqnarray}
$\xi_\phi$ is \emph{white noise}. By similar calculation, one can show the variance of $\xi_\pi$
vanishes. Thus $\xi_\pi$ can be neglected. Finally redefining $\xi=2\pi\xi_\phi/H^{3/2}$, we obtain the following equations of motion:
\begin{numcases}
	\displaystyle
	\dot{\phi}_\mathrm{IR}=\pi_\mathrm{IR}+\frac{H^{3/2}}{2\pi}\xi, \\
	 \displaystyle
	\dot{\pi}_\mathrm{IR}=-3H\pi_\mathrm{IR}-V^\prime(\phi_\mathrm{IR}), \\
	\displaystyle
	\braket{\xi(t)}=0, \quad \braket{\xi(t)\xi(t^\prime)}=\delta(t-t^\prime), \label{noise condition}
\end{numcases}
Furthermore  adopting the slow-roll approximation, $\dot{\pi}_\mathrm{IR}\simeq0$,
and using the Friedmann equation, $V\simeq3M_p^2H^2$, we obtain
\begin{eqnarray}
        \dot{\phi}_\mathrm{IR}+2M_p^2H^\prime=\frac{H^{3/2}}{2\pi}\xi(t),
\label{Stochastic EoM}
\end{eqnarray}
where $M_p\approx2.4\times10^{18}\mathrm{GeV}$ is the reduced planck mass.
Therefore it is clear that the contributions of the quantum kicks to the coarse-grained field (i.e. IR mode)
are represented as stochastic white noise.

\section{Analytic calculation of curvature perturbations}\label{analytic}

In this section we develop the algorithm for calculating correlation functions of curvature perturbations of any order in the stochastic formalism.
For this purpose, an analogy between slow-rolling inflaton dynamics and stochastic process is important. 
More specifically, a concept of the first passage time in stochastic calculus is crucial 
in order to use $\delta N$ formalism. Therefore, in the first place we briefly review stochastic calculus and 
then we explore the calculation of the curvature perturbation.

\subsection{First passage time} \label{fpt}

Dynamics of a slow-rolling inflaton in the stochastic formalism corresponds to a certain type of stochastic processes. 
As we see later, when the Hubble parameter is given by a linear function of $\phi$, the corresponding stochastic process is 
a Brownian motion with a constant drift.
Thus, we describe it in this subsection.
A Brownian motion with a drift is described by
\begin{eqnarray}
  X(t)=\mu t + W(t),
\label{Brownian motion}
\end{eqnarray}
where $X(t)$ is the position of a realization of the stochastic process at a time $t$, $\mu$ is a positive constant which represents 
a drift velocity and $W(t)$ is a Brownian motion with the initial condition $W(t=0)=0$. A stochastic property of the Brownian motion is 
given by a normal distribution whose expected value and variance are,
\begin{eqnarray}
  \braket{W(t)}=0, \quad \braket{W^2(t)}=t.
 \label{Normal distribution}
\end{eqnarray}
Note that the Brownian motion can be represented as a time integrated white noise. 
Indeed, for the white noise $\xi(t)$ normalized as eq. (\ref{noise condition}), it can be checked easily that the Brownian motion $W(t)$ defined as
$W(t)=\int^t_0\xi(t^\prime)dt^\prime$ satisfies eq. (\ref{Normal distribution}).

It is known that considering following quantity is useful to see
characteristics of a stochastic process,
\begin{eqnarray}
  Z(t)=\exp\left[\sigma X(t)-\left(\sigma\mu+\frac{1}{2}\sigma^2\right)t\right],
\end{eqnarray}
where $\sigma$ is an arbitrary positive costant. This quantity is called \emph{exponential martingale}. Provided that 
$X(t)$ is given by eq.~(\ref{Brownian motion}) and
its initial condition is $X(0)=0$, the expected value of $Z(t)$ can be computed as
\begin{eqnarray}\label{expmartingale}
  \braket{Z(t)}&=&\Braket{\exp\left(\sigma W(t)-\frac{1}{2}\sigma^2t\right)} \nonumber \\
  &=&\exp\left(-\frac{1}{2}\sigma^2t\right)
        \frac{1}{\sqrt{2\pi t}}\int^{\infty}_{-\infty} dx\, 
        \exp\left(\sigma x-\frac{x^2}{2t}\right)\notag\\
  &=&1.
\end{eqnarray}
where we have used the fact that the probability distribution of $W(t)$ is the normal distribution eq.~(\ref{Normal distribution}).
Therefore, if $X(0)=0$, the expectation value of $Z(t)$ is always 1 irrespective of $t$.

Next, we introduce the first passage time $\tau_m$. $\tau_m$ is defined as a time when a given realization of a stochastic process 
reaches $X(t)=m$ for its first time. 
Note that it is not an universal time but defined for each realization (called \emph{sample path}) 
and it takes a different value for each sample path.
In terms of inflation, it is nothing but the e-folding number $N$ because $N$ is the time which each separate universe
takes from some initial flat slice to the end of inflation. Therefore the stochastic property of $\tau_m$
is crucial for the $\delta N$ formalism. To derive it, we substitute $t= \tau_m$ into eq.~(\ref{expmartingale})\footnote{It should be
noted that $\tau_m$ depends on each sample path 
while $t$ is independent of paths and an universal time. Therefore 
the validity of this substitution looks non-trivial. However \emph{Doob's optional sampling theorem}~\cite{Doob} guarantees
its validity. One can also show that the probability of $\tau_m\to\infty$ converges to $0$~\cite{Shreve}.}
and obtain
\begin{eqnarray}
  \Braket{\exp\left[\sigma m-\left(\sigma\mu+\frac{1}{2}\sigma^2\right)\tau_m\right]}=1.
\end{eqnarray}
Defining a dummy parameter $J=\sigma\mu+\frac{1}{2}\sigma^2$, it reads
\begin{eqnarray}
  \braket{e^{-J\tau_m}}=e^{m\mu-m\sqrt{2J+\mu^2}}.
\end{eqnarray}
It is a generating function of $\langle \tau_m^n \rangle$. By differentiating it $n$ times with respective to $J$ and taking $J\to 0$, one can obtain,
\begin{eqnarray}
  \braket{\tau_m}=\frac{m}{\mu}, \quad
  \braket{\tau_m^2}=\frac{m^2}{\mu^2}+\frac{m}{\mu^3},\quad
   \braket{\tau_m^3}=\frac{m^3}{\mu^3}+3\frac{m^2}{\mu^4}+3\frac{m}{\mu^5}.
\end{eqnarray}
Note that $\braket{\tau_m}$ coincides with the result without the noise term ($\dot{X}=\mu$).
Furthermore we can also calculate, $\braket{\delta\tau_m^n}$, an expectation value of a n-th power fluctuation of $\tau_m$,
\begin{eqnarray}\label{perturbations}
  \braket{\delta\tau_m}=0,\quad
  \braket{\delta\tau_m^2}=\frac{m}{\mu^3}=\frac{1}{\mu^2}\braket{\tau_m},\quad
  \braket{\delta\tau_m^3}=3\frac{m}{\mu^5}=\frac{3}{\mu^4}\braket{\tau_m},
\end{eqnarray}
where $\delta\tau_m = \tau_m - \braket{\tau_m}$.
Calculations for higher order perturbations are straightforward.

\subsection{Power spectrum}

In this subsection, we confirm the correspondence between the dynamics of a slow-rolling inflaton and the Brownian motion with the drift. 
Then we show the $n$-point correlation functions of the curvature perturbations can be obtained by connecting the $\delta N$ formalism 
with the result of the previous subsection.
First of all, we assume that the Hubble parameter during inflation can be approximated by a linear function of $\phi$ as
\begin{eqnarray}
  H(\phi)\simeq H_0-\alpha\phi,
\label{Linear H}
\end{eqnarray}
where $\alpha$ is supposed to be a positive constant for clarity. 
From eqs.~(\ref{Linear H}) and (\ref{Stochastic EoM}), 
we obtain the equation of motion in the stochastic formalism,
\begin{eqnarray}
  \dot{\phi}(t)
  &\simeq&2M_p^2\alpha+\frac{H_0^{3/2}}{2\pi}\xi(t),
\label{Linear EoM}
\end{eqnarray}
where $\xi(t)$ is white noise whose variance is 1,
\begin{eqnarray}
  \braket{\xi(t)}=0,\quad\braket{\xi(t)\xi(t^\prime)}=\delta(t-t^\prime).
\end{eqnarray}
Here, we regard the Hubble parameter as a constant by neglecting $-\alpha\phi$ in $H(\phi)$, except only for
$-2M_p^2H^\prime$. This makes the coefficient of the noise term constant
and hence the coupling between the noise and the inflaton field is neglected.
As we will see later, the power spectrum calculated in this subsection correspond with that of the standard linear
perturbation theory exactly in these approximations.

Using $N=H_0t$, eq.~(\ref{Linear EoM}) is rewritten as
\begin{eqnarray}
  \frac{2\pi}{H_0}\dif{\phi}{N}(N)\simeq\frac{4\pi M_p^2\alpha}{H_0^2}+\tilde{\xi}(N),
\label{EoM in N}
\end{eqnarray}
$\tilde{\xi}(N)$ is also normalized white noise ($\tilde{\xi}(N)=\xi(t)/H_0^{1/2}$). If we define 
\begin{equation}
X(N)=\frac{2\pi\phi}{H_0},\quad \mu=\frac{4\pi M_p^2\alpha}{H_0^2},
\label{X mu correspondence}
\end{equation}
the correspondence between the inflaton dynamics eq.~(\ref{EoM in N}) and the Brownian motion with the drift eq.~(\ref{Brownian motion}) becomes 
transparent because integrated white noise is a Brownian motion as mentioned in subsection \ref{fpt}.
In brief, when eq.~(\ref{EoM in N}) is integrated by $N$, it reads
\begin{eqnarray}
	X(N)-X_i=\mu N+W(N),
\end{eqnarray}
with $X_i$ denoting the initial value of $X(N)$.

Let us assume that slow-roll inflation ends when $\phi$ reaches $\phi_f$. Then, according to the $\delta N$ formalism, 
the fluctuation of the e-folding number $N$ defined between an initial flat slice and the $\phi=\phi_f$ slice is nothing but 
the curvature perturbations in real space. Here, 
since $N$ in the stochastic dynamics corresponds to the first
passage time $\tau_m$ of the Brownian motion with the drift,
one can find that the correlators of $\delta N$ can be computed in the exactly same way as $\braket{\delta\tau_m^n}$.
Then we write the generating function,
\begin{eqnarray}
        \braket{e^{-JN}}=e^{(X_f-X_i)(\mu-\sqrt{2J+\mu^2})},
\end{eqnarray}
where $X_f$ is the final value of $X=2\pi\phi/H_0$.
Therefore we can calculate $\braket{\delta N^2}$ similarly to eq. (\ref{perturbations}) as
\begin{eqnarray}
  \braket{\delta N^2} = \braket{N^2}-\braket{N}^2 =\frac{1}{\mu^2}\braket{N}.
\end{eqnarray}
On the other hand, a power spectrum of $\delta N$
is defined by a 2-point correlator of $N$,
\begin{eqnarray}
  \mathcal{P}_{\delta N}(k) =\frac{k^3}{2\pi^2}\int d^3x\braket{\delta N(0)\delta N(x)}e^{-i\mathbf{k}\cdot\mathbf{x}}.
\end{eqnarray}
Provided that only modes which exits the horizon during inflation contribute to the variance of $N$, one can show
\begin{eqnarray}
  \braket{\delta N^2}=\int^{k_f}_{k_i}\mathcal{P}_{\delta N}(k^\prime)\frac{dk^\prime}{k^\prime}
  =\int^{\ln k_f}_{\ln k_f-\braket{N}}\mathcal{P}_{\delta N}dN.
\end{eqnarray}
Here, $k_f=a_fH_0,k_i=a_iH_0=k_fe^{-\braket{N}}$ and subscript $i$ and $f$ represent some initial slice and
the end of inflation respectively. Therefore
we obtain
\begin{eqnarray}
  \mathcal{P}_\zeta(k)=\mathcal{P}_{\delta N}(k)&=&\left.\dif{}{\braket{N}}\braket{\delta N^2}
  \right|_{\braket{N}=\ln (k_f/k)} = \frac{1}{\mu^2} = \left(\frac{H_0^2}{4\pi M_p^2\alpha}\right)^2.
\end{eqnarray}
Using a slow-roll parameter $\epsilon_H=2M_p^2(H^\prime/H)^2\simeq2M_p^2\alpha^2/H_0^2$ and
the Friedmann equation $V\simeq3M_p^2H_0^2$, we obtain the exactly same result
as the standard linear perturbation theory,
\begin{eqnarray}\label{spectrum}
  \mathcal{P}_\zeta=\frac{1}{24\pi^2M_p^4}\frac{V}{\epsilon_H}.
\end{eqnarray}
It is the main result of this paper.

What does our main result mean?
Even in more general cases where the Hubble parameter can not be approximated linearly, 
as long as inflation is a single field slow-roll type, we can obtain 
the same result by dividing the whole inflation periods into shorter periods during which the
linear approximation of the Hubble makes sense. In this case, the right hand side of eq. (\ref{spectrum}) will be evaluated
at the horizon exit of the scale of the interest because the Hubble parameter $H$ and its gradient $\alpha$ vary over 
each period.
That is because the super-horizon scale curvature perturbations are conserved in single field inflation models. Therefore,
we do not have to wait until the end of inflation and can take the final slice of the $\delta N$ formalism a certain time
when the scale of the interest becomes sufficiently larger than the horizon.
Of course, in order to use the $\delta N$ formalism, it is needed  that 
the end of the period can be taken well after the horizon exit of the interest mode, However as long as the
slow-roll approximation is good enough (i.e. the Hubble parameter does not vary rapidly), it seems possible to take such a period.
Therefore it is suggested that 
the power spectrum of the curvature perturbations will correspond with the standard result even if the stochastic effect is considered in
single slow-roll inflation.

It should be noted that in our formalism, the condition that the quantum kicks are smaller than the classical motion is
not required at all because we do not expand the inflaton field perturbatively. In general, it is said that inflation is \emph{stochastic}
if the quantum kicks are comparable to or bigger than the classical motion. 
Since $\mu$ in eq.~(\ref{X mu correspondence}) represents the ratio of the classical motion to the quantum kicks, 
inflation is stochastic for $\mu^{-1}(=\mathcal{P}_\zeta^{1/2})\gtsim1$.\footnote{Since $\mathcal{P}_\zeta$ should be $\ltsim10^{-2}$ 
due to the PBH constraints~\cite{Carr:2009jm}, we do not have to consider very large curvature perturbations.}
Apparently it seems that the larger the power spectrum became
(i.e. the more stochastic inflation becomes), the worse the linear perturbation theory is. However our main result implies that
even if inflation is stochastic to a certain extent, the linear perturbation theory gives a precise result for single field slow-roll inflation.
It is because in single field slow-roll inflation the non-Gaussianity of the curvature perturbations is small even if the power spectrum
is large. In the $\delta N$ formalism, when the non-Gaussianity is small, the first order approximation,
\begin{eqnarray}
        \zeta=\delta N\simeq\pdif{N}{\phi}\delta\phi,
\end{eqnarray}
is good approximation regardless of the amplitude of $\delta N$ and it is nothing but the linear perturbation theory.
Our result not only verifies our formalism but also guarantees that 
the linear perturbation theory gives the precise curvature perturbations in single field slow-roll inflation even if the quantum fluctuations are 
comparable to classical dynamics.

\subsection{Non-Gaussianity}

As we see in section \ref{fpt}, we can calculate higher order expected values of the e-foldings easily in our formalism.
Therefore we can also obtain the non-Gaussianity of the curvature perturbations.
One can expect that the curvature perturbations are almost Gaussian 
because we consider single field slow-roll inflation models.
Let us check it.
For this purpose, the quantities such as the skewness which vanishes
in an exact Gaussian case is useful. The skewness $S_\zeta$ is defined by $S_\zeta:=\braket{\zeta^3}
/\braket{\zeta^2}^{3/2}$ and in our case it is given by,
\begin{eqnarray}
  S_\zeta=\frac{\braket{\zeta^3}}{\braket{\zeta^2}^{3/2}}=\frac{3}{\mu}\frac{1}{\braket{N}^{1/2}}
  =\mathcal{P}_\zeta^{1/2}\frac{3}{\braket{N}^{1/2}}.
\end{eqnarray}
Here we have used the result of section \ref{fpt} for $\braket{\zeta^3}$. 
In the observational aspects, the super-horizon modes for the current observable universe can not
be distinguished from the homogeneous zero mode. We can observe only the sub-horizon modes corresponding
to $k_f>k>k_fe^{-N_\mathrm{obs}}$ where $N_\mathrm{obs}\sim60$. Therefore the expected values should be
averaged over the current observable universe, and then we take $\braket{N}\sim60$.
If the power spectrum is consistent with the observed value $\mathcal{P}_\zeta^{1/2}\sim10^{-5}$,
the skewness is indeed much less than unity and then curvature perturbations are indeed almost Gaussian. 

The commonly used nonlinearity parameter $f_\mathrm{NL}$ is also useful. Let us calculate the
local type nonlinearity parameter by approximating curvature perturbations as
\begin{eqnarray}\label{definition of fnl}
  \zeta(\mathbf{x})\simeq g(\mathbf{x})+\frac{3}{5}f_\mathrm{NL}^\mathrm{loc}
  (g^2(\mathbf{x})-\braket{g^2(\mathbf{x})}),
\end{eqnarray}
where $g(\mathbf{x})$ is a Gaussian fluctuation. Since curvature perturbations are almost Gaussian,
$f_\mathrm{NL}^\mathrm{loc}$ is expected to be much less than unity.
Now the squared and cubic expected values are
\begin{eqnarray}
  \braket{\zeta^2}&=&\braket{g^2}+\mathcal{O}(f_\mathrm{NL}^\mathrm{loc\,2}), \\
  \braket{\zeta^3}&=&\frac{9}{5}f_\mathrm{NL}^\mathrm{loc}(\braket{g^4}-\braket{g^2}^2)
  +\mathcal{O}(f_\mathrm{NL}^\mathrm{loc\,3}) 
  \simeq \frac{18}{5}f_\mathrm{NL}^\mathrm{loc}\braket{g^2}^2,
\end{eqnarray}
where we used the Wick's theorem $\braket{g^4}=3\braket{g^2}^2$. Therefore  at leading order we obtain
\begin{eqnarray}
  f_\mathrm{NL}^\mathrm{loc}&\simeq&\frac{5}{18}\frac{\braket{\zeta^3}}{\braket{\zeta^2}^2}
  =\frac{5}{6}\frac{1}{\braket{N}}\sim\frac{1}{72}.
\label{Stochastic fnl}
\end{eqnarray}
For the same reason we mentioned above, we take $\braket{N}\sim60$.  

For comparison with it, let us calculate the nonlinearity parameter by the standard linear perturbation theory, too.
In the $\delta N$ formalism, taking the second order perturbations of the inflaton field,
\begin{eqnarray}
        \zeta(\mathbf{x})=\delta N(\mathbf{x})\simeq\pdif{N}{\phi}\delta\phi(\mathbf{x})
        +\frac{1}{2}\frac{\partial^2N}{\partial\phi^2}\delta\phi^2(\mathbf{x}).
\end{eqnarray}  
Because the inflaton is assumed to be almost massless, $\delta\phi$ is a nearly Gaussian fluctuation and we can assume $g(\mathbf{x})=N^\prime\delta\phi(\mathbf{x})$.
Therefore, comparing with eq. (\ref{definition of fnl}), we obtain
\begin{eqnarray}
        \frac{3}{5}f_\mathrm{NL}^\mathrm{loc}=\frac{1}{2}\frac{N^{\prime\prime}}{(N^\prime)^2}.
\end{eqnarray}
The equation of motion without noise is
\begin{eqnarray}
        H\dif{\phi}{N}&=&-2M_p^2H^\prime, 
\end{eqnarray}
By integrating it, one can find the e-foldings number and its derivatives,
\begin{eqnarray}\label{efoldings}
        N=\frac{1}{2M_p^2\alpha}\int^{\phi_f}_\phi(H_0-\alpha\phi^\prime)d\phi^\prime,\quad
        N^\prime \simeq -\frac{H_0}{2M_p^2\alpha},\quad
        N^{\prime\prime}=\frac{1}{2M_p^2}.
\end{eqnarray}
Then the nonlinearity parameter is given by
\begin{eqnarray}
        f_\mathrm{NL}^\mathrm{loc}=\frac{5}{6}\times\frac{2M_p^2\alpha^2}{H_0^2}=\frac{5}{6}\epsilon_H.
\end{eqnarray}
Although it is apparently different from the stochastic result eq. (\ref{Stochastic fnl}), they are consistent actually.
With the definition of the slow-roll parameter $\epsilon_H=2M_p^2(H^\prime/H)^2$ and the slow-roll equation of motion $\dot{\phi}+2M_p^2H^\prime=0$,
we rewrite the slow-roll parameter
\begin{eqnarray}
	\epsilon_H=-\frac{\dot{H}}{H^2}.
\end{eqnarray}
It can be deformed as
\begin{eqnarray}
	Hdt=-\frac{1}{\epsilon_H}\frac{dH}{H}.
\end{eqnarray}
Then supposing the time dependence of the slow-roll parameter can be neglected, it reads
\begin{eqnarray}
	\int^{t_f}_{t_i}Hdt=\frac{1}{\epsilon_H}\log\left(\frac{H_i}{H_f}\right).
\end{eqnarray}
The left hand side is nothing but the precise definition of the e-foldings $N$.
In slow-roll inflation, $\log(H_i/H_f)$ is expected to be an order one factor, so it has been proved that the e-foldings $N$ and
the inverse of the slow-roll parameter $\epsilon_H^{-1}$ have the same order of magnitude.
As a result, the nonlinearity parameter in the stochastic formalism is consistent with that obtained in the standard linear perturbation theory.

\section{Extension to more general case}\label{numerical}

Though we explain the analytic calculation of the power spectrum and the non-Gaussianity
in single field slow-roll inflation in the
last section, we have to solve the equation of motion numerically in general cases. 
In this section, we briefly describe the algorithm for numerical calculations.
For single field inflation models,
\begin{itemize}
\item[1.]
  Choose a certain initial value $\phi_i$ for the inflaton field $\phi$ temporarily.
\item[2.]
  Start inflation from that temporary initial value $\phi_i$, integrate the full
  equations of motion,
        \begin{numcases}
                        \displaystyle
                        H\dif{\phi}{N}=\pi+\frac{H^2}{2\pi}\xi(N),  \label{eom of phi} \\
                        \displaystyle
                        H\dif{\pi}{N}=-3H\pi-\dif{V}{\phi}, \label{eom of pi}
        \end{numcases}
  numerically and calculate the e-foldings $N$ which is needed to reach the
  end value $\phi_f$ (for example, the point where the slow-roll parameter becomes unity).
  In the numerical integration of eqs.~(\ref{eom of phi}) and (\ref{eom of pi}), the noise term $\xi(N)$ is replaced with a random variable obeying a normal
  distribution whose expected value and variance are 0 and $\Delta N$ where $\Delta N$ is the step of the
  numerical integration. Therefore, the calculated e-folding varies in each calculation (realization).
  Repeating this calculation many times, we can get $\braket{N}$, $\braket{\delta N^2}$, $\braket{\delta N^3}$
  and so on.
\item[3.]
  Reiterating the above process for various initial values, we obtain the relation between
  $\braket{N}$ and $\braket{\delta N^2}$ (or $\braket{\delta N^3}$), and thus $\braket{\delta N^2}$ 
  and $\braket{\delta N^3}$ can be represented as functions of $\braket{N}$.
\item[4.]
  Finally, differentiating $\braket{\delta N^2}$ by $\braket{N}$, we can obtain the
  power spectrum of curvature perturbations,
  \begin{eqnarray}
    \mathcal{P}_\zeta(k)=\left.\dif{\braket{\delta N^2}}{\braket{N}}\right|_{\braket{N}=\ln k_f-\ln k}.
  \end{eqnarray}
  We can also obtain the local type nonlinearity parameter $f_\mathrm{NL}^\mathrm{loc}$ when the curvature perturbations are almost Gaussian,
  \begin{eqnarray}
    f_\mathrm{NL}^\mathrm{loc}\simeq\left.\frac{\braket{\delta N^3}}{\braket{\delta N^2}^2}\right|_{\braket{N}\sim60}.
  \end{eqnarray}
        If the curvature perturbations are not Gaussian at all, one should use the full simultaneous equations,
        \begin{numcases}
                        \displaystyle
                        \braket{\zeta^2}=\braket{g^2}+\frac{18}{25}f_\mathrm{NL}^\mathrm{loc\,2}\braket{g^2}^2, \\
                        \displaystyle
                        \braket{\zeta^3}=\frac{18}{5}f_\mathrm{NL}^\mathrm{loc}\braket{g^2}^2
                        +\frac{216}{125}f_\mathrm{NL}^\mathrm{loc\,3}\braket{g^2}^3,
        \end{numcases}
        for $f_\mathrm{NL}^\mathrm{loc}$ and $\braket{g^2}$.
\end{itemize}

The extension to multi-field inflation models is also interesting. In a multi-field case, however,
the choice of the temporary initial field values is not trivial. It is because
the solution of the equation of motion with noise (called sample path) can not be
determined uniquely because the field space is multi-dimensional
and can be different from the no noise solution (called classical path).
Thus a careful treatment is necessary. We propose the following procedure.
\begin{itemize}
\item[1.]
	Choose some fixed initial values for relevant fields. These initial values must be placed at regions where the inflaton potential is high enough
	for inflation to last at least 60 e-foldings.
\item[2.]
        Produce a lot of sample path from that fixed initial value. Each sample path is assumed to have a same probabilistic weight.
        It is the assumption of stochastic calculus. 
\item[3.]
        For each sample path, take various temporary initial values on the sample path and calculate 
        the power spectrum similarly to single field inflation.
\item[4.]
        The true power spectrum should be identified with the mean of those obtained from many sample paths.
        In other words, since each sample path has same probabilistic weight, 
        sum up the power spectrum calculated for each sample path and divide it by the number of sample paths. 
\end{itemize}
In this algorithm, all sample paths are produced from one fixed initial values. It means that in the sufficiently early phase of inflation,
the inflaton field values were same at least over the observable universe, and thereafter each horizon evolves separately.
It is valid because the observable universe was in one same horizon more than 60 e-foldings before the end of inflation
and the assumption that the inflaton field values were same at that time is natural. 
There is a problem concerning the fixed initial values.
Generally, if there is more than one scalar fields whose masses are much lighter than the Hubble parameter, 
the paths followed by the inflaton fields quite depend on their initial values.
Thus, different choices of the initial values may result in different power spectra.
However, in many cases, such as the hybrid inflation model, the inflaton field value will converge to a classical attractor path soon
without dependence on the initial values. Therefore, one can take the fixed initial values on that classical attractor in such cases.

\section{Conclusion}\label{Conclusion}
In this paper, we propose a new algorithm for calculating the curvature perturbations 
produced during inflation in the stochastic formalism. In our algorithm, the curvature perturbations can be 
calculated non-perturbatively. As mentioned in section \ref{analytic}, in single field slow-roll inflation,
we can obtain the curvature perturbations analytically in our formalism with the help of the techniques of stochastic calculus,
and the calculated power spectrum and non-Gaussianity of the curvature perturbations
are consistent with the standard result.
It is quite natural because the curvature perturbations produced in single slow-roll inflation are
almost Gaussian and the linear perturbation approach (i.e. the standard approach) works well.
Therefore, the power spectrum is not expected to be so different from linear perturbation
result even if the stochastic effect is considered.
In more general cases, such as highly non-Gaussian cases, our formalism is expected to give more precise result than 
the standard linear approach. Therefore, we also described how the curvature perturbations can be calculated
by our formalism in such cases in section \ref{numerical}.
Even though it is known that the non-Gaussianity is so small at CMB scale, it may not be negligible
at smaller scales. Furthermore, the power spectrum at small scales may also be large and some interesting
objects like PBHs and UCMHs can be formed. 
Our non-perturbative formalism is useful for studying such cases.


\acknowledgments

We would like to thank Shuichiro Yokoyama for helpful discussions. 
This work is supported by Grant-in-Aid for Scientific research from the Ministry of Education, Science, Sports, and Culture (MEXT), Japan, No. 25400248
[MK], No. 21111006 [MK] and also by World Premier International Research Center Initiative (WPI Initiative), MEXT, Japan.
T.F. and T.T. acknowledge the support by JSPS Research Fellowships for Young
Scientists.The work of Y.T. is partially supported by an Advanced Leading Graduate Course for Photon Science grant.


\end{document}